\documentclass[aps,prl,twocolumn,showpacs,superscriptaddress]{revtex4}
\usepackage{graphicx}

\begin{document}

\title{Normal-state hourglass dispersion in spin excitations of FeSe$_{x}$Te$_{1-x}$}

\author{Shiliang Li}
\affiliation{Beijing National Laboratory for Condensed Matter Physics, Institute of Physics, Chinese Academy of Sciences, Beijing 100190, China
}
\email{slli@aphy.iphy.ac.cn}
\author{Chenglin Zhang}
\affiliation{
Department of Physics and Astronomy, The University of Tennessee, Knoxville, Tennessee 37996-1200, USA
}
\author{Meng Wang}
\affiliation{Beijing National Laboratory for Condensed Matter Physics, Institute of Physics, Chinese Academy of Sciences, Beijing 100190, China
}
\affiliation{
Department of Physics and Astronomy, The University of Tennessee, Knoxville, Tennessee 37996-1200, USA
}
\author{Hui-qian Luo}
\affiliation{Beijing National Laboratory for Condensed Matter Physics, Institute of Physics, Chinese Academy of Sciences, Beijing 100190, China
}
\author{Xingye Lu}
\affiliation{Beijing National Laboratory for Condensed Matter Physics, Institute of Physics, Chinese Academy of Sciences, Beijing 100190, China
}
\author{Enrico Faulhaber}
\affiliation{
Gemeinsame Forschergruppe HZB - TU Dresden, D-14109 Berlin, Germany
}
\affiliation{
Forschungsneutronenquelle Heinz Maier-Leibnitz (FRM-II), TU M$\ddot{u}$nchen, D-85747 Garching, Germany}
\author{Astrid Schneidewind}
\affiliation{
Gemeinsame Forschergruppe HZB - TU Dresden, D-14109 Berlin, Germany
}
\affiliation{
Forschungsneutronenquelle Heinz Maier-Leibnitz (FRM-II), TU M$\ddot{u}$nchen, D-85747 Garching, Germany}
\author{Peter Link}
\affiliation{
Forschungsneutronenquelle Heinz Maier-Leibnitz (FRM-II), TU M$\ddot{u}$nchen, D-85747 Garching, Germany}
\author{Jiangping Hu}
\affiliation{
Department of Physics, Purdue University, West Lafayette, IN 47907, USA}
\affiliation{Beijing National Laboratory for Condensed Matter Physics, Institute of Physics, Chinese Academy of Sciences, Beijing 100190, China
}
\author{Tao Xiang}
\affiliation{Beijing National Laboratory for Condensed Matter Physics, Institute of Physics, Chinese Academy of Sciences, Beijing 100190, China
}
\affiliation{Institute of Theoretical Physics, Chinese Academy of Sciences, P.O. Box 2735, Beijing 100190, China}
\author{Pengcheng Dai}
\email{daip@ornl.gov}
\affiliation{
Department of Physics and Astronomy, The University of Tennessee, Knoxville, Tennessee 37996-1200, USA
}
\affiliation{
Neutron Scattering Science Division, Oak Ridge National Laboratory, Oak Ridge, Tennessee 37831-6393, USA
}
\affiliation{Beijing National Laboratory for Condensed Matter Physics, Institute of Physics, Chinese Academy of Sciences, Beijing 100190, China
}
\begin{abstract}
We use cold neutron spectroscopy to study the low-energy spin excitations of superconducting (SC) FeSe$_{0.4}$Te$_{0.6}$ and essentially non-superconducting (NSC) FeSe$_{0.45}$Te$_{0.55}$. In contrast to
BaFe$_{2-x}$(Co,Ni)$_{x}$As$_2$, where the low-energy spin excitations are commensurate both in the SC and normal state, the normal-state spin excitations in SC FeSe$_{0.4}$Te$_{0.6}$ are incommensurate and show an hourglass dispersion near the resonance energy. Since similar hourglass dispersion is also found in the NSC FeSe$_{0.45}$Te$_{0.55}$, we argue that the observed incommensurate spin excitations in FeSe$_{1-x}$Te$_{x}$ are not directly associated with superconductivity. Instead, the results can be understood within a picture of Fermi surface nesting assuming extremely low Fermi velocities and spin-orbital coupling.
\end{abstract}



\maketitle
The discovery of antiferromagnetism in the Fe-based high-transition (high-$T_c$) temperature superconductors \cite{kamihara,rotter,mawkuen2,fangmh,bao,slli} has reinvigorated research in understanding the role of magnetism in the superconductivity \cite{mazin,chubukov,fwang,seo}.  Among the different classes of iron pnctides,
FeSe$_{x}$Te$_{1-x}$ has the simplest crystal structure, composed of
layers of Fe atoms forming a square lattice with Se/Te atoms
centered above or below these squares alternating in
a checkerboard fashion (Fig. 1a) \cite{mawkuen2,fangmh,bao,slli}. To obtain a comprehensive understanding for the role of magnetism in the superconductivity of FeSe$_{x}$Te$_{1-x}$ , it is important to determine the energy ($E$) and momentum ($Q$) dependence of its spin excitations in the normal and superconducting (SC) state, and compare the results with that of the other Fe-based \cite{christian,lumsden,chi,slli2,inosov,clester} and cuprate high-$T_c$ superconductors \cite{vignolle,hayden}.
For cuprates such as La$_{2-x}$Sr$_x$CuO$_4$ \cite{vignolle} and YBa$_2$Cu$_3$O$_{6+x}$ \cite{hayden}, the low-energy spin excitations consist of a spin gap, a quartet of incommensurate peaks merging into a collective excitation called the resonance, and then dispersing outward again at energies above the mode forming a hourglass-like dispersion. If the resonance in cuprates arises from metallic nested Fermi surfaces,
the incommensurate scattering below the resonance signal the strong anisotropy of the $d$-wave SC gap \cite{norman,eschrig}. In contrast, the experimental observation of a commensurate resonance above a clean spin gap for the Fe-based BaFe$_{2-x}$(Co,Ni)$_x$As$_2$ superconductors \cite{lumsden,chi,slli2,inosov,clester} has been taken as evidence that the mode here arises from quasiparticle excitations from sign reversed isotropic $s$-wave SC gaps on the hole and electron Fermi pockets (Figs. 1c-1e) \cite{mazin,maier1,korshunov}. If Fermi surfaces in FeSe$_{x}$Te$_{1-x}$ are
similar to that of BaFe$_{2-x}$(Co,Ni)$_x$As$_2$ \cite{subedi}, one should expect low energy
spin excitations to behave similarly as well.  Indeed, the observation of a commensurate
neutron spin resonance in FeSe$_{x}$Te$_{1-x}$ at $x=0.4,0.5$ \cite{yqiu,mook} appears to confirm this conclusion. Although recent  neutron scattering
measurements have found incommensurate peaks (Fig. 1b) at energies near and above the resonance energy \cite{lumsden1,argyriou,lee,zxu}, these results
are consistent with the current theory \cite{argyriou} and incommensurability has been
interpreted as the signature of spin-orbital correlations \cite{lee}. If the SC gaps in FeSe$_{x}$Te$_{1-x}$ for $x\sim 0.5$
are indeed isotropic $s$-wave, the resonance should not show an hourglass-like dispersion
as in the case of cuprates \cite{argyriou}.

Surprisingly, we found that spin excitations in SC FeSe$_{0.4}$Te$_{0.6}$ have an hourglass-like dispersion (Fig. 1f)
in the normal state. On cooling below $T_c$, the excitation spectrum opens a small
spin gap before merging into the resonance. In the nearly non-superconducting (NSC) FeSe$_{0.45}$Te$_{0.55}$ that shows weak resonance and no spin gap, similar hourglass dispersion was also found. 
Therefore, such dispersion must be unrelated to the SC electronic gap, which is contrary to the case of cuprate superconductors \cite{vignolle,hayden,norman,eschrig}.  Instead, we argue that the results are similar to
that for pure chromium \cite{endoh,fishman} and can be understood by the multi-band nature of the system.

\begin{figure}
\includegraphics[scale=.70]{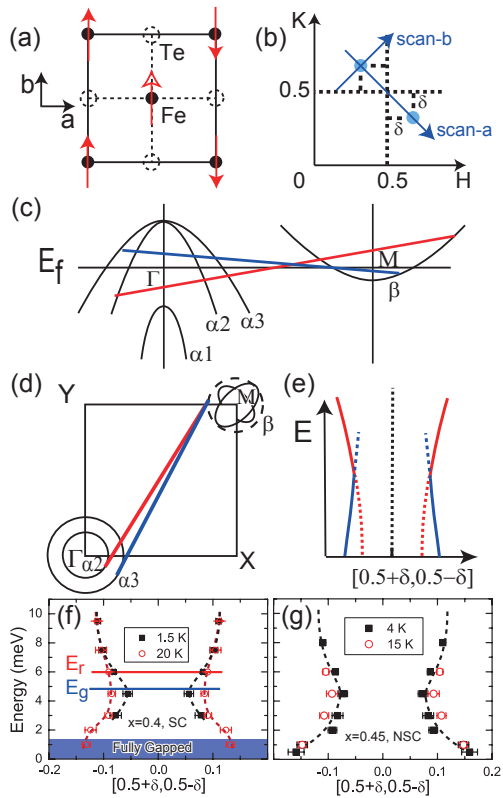}
\caption{ (Color online) (a)
Schematic in-plane spin structure of the nonsuperconducting
FeTe, where the solid and hollow arrows
represent two sublattices of spins which can be either parallel or
antiparallel \cite{slli}.
Upon substitution of Se for Te to form FeSe$_{x}$Te$_{1-x}$, the static long-range AF order is suppressed,
and the system display strong spin excitations at incommensurate positions near $Q=(0.5,0.5)$ as shown in (b).
The incommensurate scattering only appear at positions $(0.5-\delta,0.5+\delta)$ and $(0.5+\delta,0.5-\delta)$.
Our transverse scans are along the scan direction {\bf a}, and the scan along the incommensurate position
that is perpendicular to scan-{\bf a} is marked as scan-{\bf b}.
(c,d) Schematic diagram of Fermi surfaces near $\Gamma$ and $M$ points from results of
recent photoemission experiments \cite{tamai,fchen}. (e) In a multiband itinerant picture,
quasiparticle excitations from the $\alpha2$ band to the $\beta$ band can give rise to the upper branch of
the hourglass dispersion as shown in the solid red lines. The lower branch of the dispersion is then
a consequence of the excitations from the $\alpha3$ band to the $\beta$ band.
(f) Experimental determination of the spin excitation dispersions in the normal (open red circles) and SC
(black filled squares) states of FeSe$_{0.4}$Te$_{0.6}$.  A full spin gap opens below $E\approx 1$ meV at 1.5 K.
The magnitude of $E_g$ marks the energy below which intensity of spin excitations decrease below $T_c$, whereas $E_r$ indicates the resonance energy.
The dispersion curves are obtained by fitting two Gaussians on linear backgrounds through transverse scans in Figs. 2 and 3. We note that the incommensurability at 2 meV at 20 K is obtained by fitting the difference between the 20 K data and 1.5 K data. The horizontal error bars are the fitted errors of the incommensurability. (g) Hourglass dispersion in the NSC FeSe$_{0.45}$Te$_{0.55}$ at 4 K and 15 K.  
 }
\label{structure}
\end{figure}

We have carried out our inelastic neutron scattering experiments using the PANDA
cold triple-axis spectrometer at the FRM-II, TU M$\rm\ddot{u}$nchen, Germany. We used pyrolytic graphite PG$(0,0,2)$
as monochromator and analyzer without any collimator.  The final
neutron wave vector was fixed at $k_f=1.55 \rm\ \AA^{-1}$ with a
cooled Be filter before the analyzer. The energy resolution is about 0.15 meV.
We chose to study SC FeSe$_{x}$Te$_{1-x}$ with $x=0.4$ because previous measurements on similar samples
have shown the presence of the resonance at $E=6.5$ meV and suppression of spin fluctuations below $E\approx 4$ meV \cite{yqiu,mook,argyriou,lee}.
We define the momentum transfer $Q$ at ($q_{x}$,$q_{y}$,$q_{z}$) as ($H$,$K$,$L$)=($q_{x}a/2\pi $,$
q_{y}b/2\pi $,$q_{z}c/2\pi $) reciprocal lattice units (rlu), where
the lattice parameters of the tetragonal unit cell ($P4/nmm$ space group) are $a=b=3.786$ \AA\ and $c=6.061$ \AA\ (Figs. 1a and 1b).
We co-aligned $\sim$15 grams of single crystals of $x=0.4$ samples to within 1.5$^\circ$
(prepared using flux method similar to \cite{yqiu} with $T_c=14$ K)
in the $[H,K,0]$ scattering plane with $c$-axis vertical.
All momentum transfers given as $Q=(H,K)$
are read to be $Q=(H,K,0)$. To further study the effect of superconductivity, we have also measured
a poorly superconducting (NSC) FeSe$_{x}$Te$_{1-x}$ ($x=0.45$) sample ($\sim$ 23 grams) with a $T_c$ of 10 K and a superconducting volume less than 30\% in the same scattering zone.
The samples were loaded inside either a variable temperature liquid He cryostat or a closed cycle cryostat.
Since superconductivity and magnetic order in Fe$_{1+\delta}$Se$_{x}$Te$_{1-x}$
are extremely sensitive to the excess Fe content $\delta$ \cite{bao,slli,mcqueen09}, we have measured
 the excess Fe in our samples using inductively coupled plasma atomic-emission spectroscopy
analysis.   We find that both samples are essentially stoichiometric without excess Fe to within 2\% of the measurement accuracy. This is consistent with earlier results that suggest
Fe$_{1+\delta}$Se$_{x}$Te$_{1-x}$ samples in this Se-range have little excess Fe \cite{fangmh,bao,slli}.

Since magnetic scattering of $x=0.4$ sample near the resonance energy
are highly two-dimensional and peak at $(0.5-\delta,0.5+\delta)$ and $(0.5+\delta,0.5-\delta)$
positions as shown in Fig. 1b  \cite{lumsden1,argyriou,lee}, we focus our attention to the energy evolution of these
two peaks by carrying out transverse scans along the $[H,1-H,0]$ direction (scan-{\bf a} direction in Fig. 1b).
At 20 K ($T=T_c+6$ K), the scattering at $E=1$ meV show two clear peaks centered at $(0.5-\delta,0.5+\delta)$ and $(0.5+\delta,0.5-\delta)$
positions respectively with $\delta=0.132 \pm 0.007$ (Fig. 2a, open circles).  Upon cooling to 1.5 K ($T=T_c-12.5$ K),
the incommensurate peaks disappear and the scattering becomes featureless indicating the opening of a spin gap (Fig. 2a, filled squares).
To confirm that the magnetic scattering at $E=1$ meV is indeed incommensurate as depicted in Fig. 1b, we carried out  scans along the scan-{\bf b} direction. The outcome of these scans (Fig. 2b) indeed shows a peak at the expected position ($H=0$ and $\delta=0.13$) in the normal state and it disappears below $T_c$.
The temperature dependence of the scattering at $E=1$ meV and $Q=(0.5-\delta,0.5+\delta)$, where $\delta=0.13$, in Fig. 2c shows a sudden reduction in intensity near $T_c$.  For comparison, the intensity of the resonance at $E=6$ meV and $Q=(0.5,0.5)$ below $T_c$ (Fig. 2c) clearly increases below $T_c$ \cite{yqiu,mook}.

\begin{figure}[tbp]
\includegraphics[scale=.45]{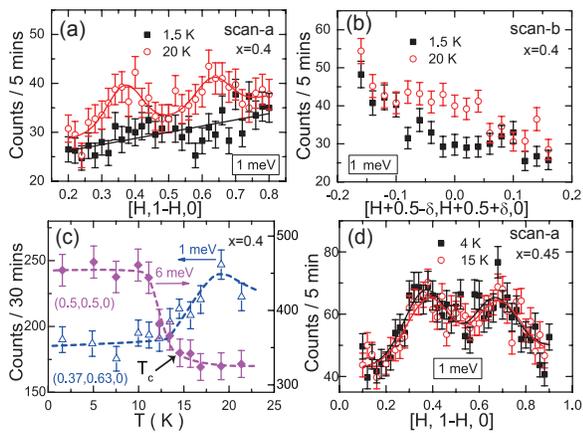}
\caption{(Color online) (a) Constant energy scans at $E=1$ meV along the scan-{\bf a} direction
below (solid squares) and above $T_c$ (open circles). The normal state incommensurate scattering
is completely suppressed below $T_c$.  The solid line is a fit of the data using two Gaussians on a sloped
linear background.  The data at 1.5 K are featureless indicating the presence of a full spin gap at this energy.
(b) Constant-energy scans along the scan-{\bf b} direction below and above $T_c$.  The data confirm that the normal state
magnetic scattering is incommensurate and centered at $(0.5-\delta,0.5+\delta)$ and $(0.5+\delta,0.5-\delta)$ with
$\delta=0.132 \pm 0.007$ at $E=1$ meV.  The scattering again become featureless below $T_c$.
(c) The temperature dependence of the scattering at $Q=(0.37,0.63)$ and $E=1$ meV (left scale) decreases below $T_c$, while
the scattering at the resonance energy [$Q=(0.5,0.5)$ and $E=6$ meV] increases below $T_c$ (right scale).
(d) Similar scans as (a) in the $x=0.45$ NSC sample.
  }
\label{phasetransition}
\end{figure}

\begin{figure}[tbp]
\includegraphics[scale=0.45]{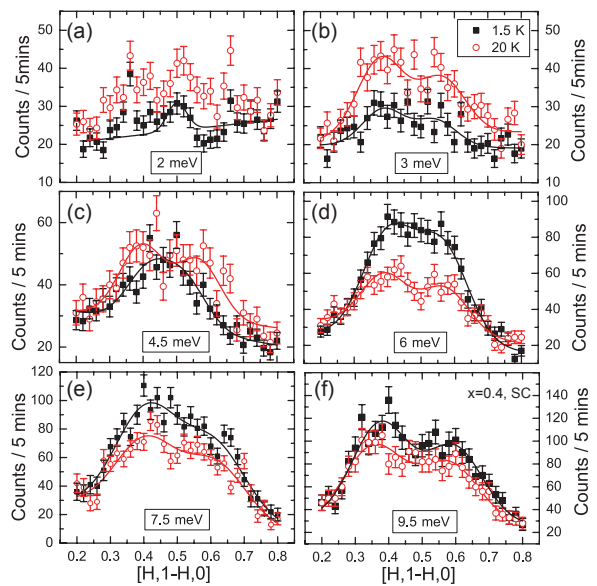}
\caption{ (Color online) Constant-energy scans along the scan-{\bf a} direction above and below $T_c$
for energies (a) 2 meV; Here the transverse scans have three peaks in the normal state, centered at $(0.5-\delta,0.5+\delta)$, $(0.5,0.5)$,
and $(0.5+\delta,0.5-\delta)$, and these peaks are suppressed but not eliminated below $T_c$.
(b) 3 meV; (c) 4.5 meV; (d) 6 meV; (e) 7.5 meV; and (f) 9.5 meV.
The solid lines are Gaussian fits to the data on linear sloped backgrounds.  Superconductivity has opposite effect on
spin excitations below and above $E=4.5$ meV.
 }
\label{doping}
\end{figure}

Fig. 3 summarizes the energy dependence of the transverse scans at a series of energies below and above $T_c$ for the $x=0.4$ sample.  The scattering is incommensurate at all energies investigated but display quite different temperature dependence for spin excitation energies below and above $E=4.5$ meV.
For energies between $E=2$ and 4.5 meV, the scattering in the normal state is suppressed but not eliminated at 1.5 K (Figs. 3a-3c). The effect of superconductivity reduces the entire transverse scattering profile at $E=3$ meV (Fig. 3b),
but only suppresses the incommensurate magnetic scattering intensity at $E=4.5$ meV (Fig. 3c). The $E=2$ meV excitations in both the normal and superconducting state are difficult to fit, although the suppression of the signal is still clear. For energies of $E=6$, 7.5, and 9.5 meV,
the low-temperature scattering are strongly enhanced due to the resonance. All these data were fitted by two Gaussian functions with equal width as shown by the solid lines in Fig. 3a-3f, where the incommensurability $\delta$ is defined as half of the distance between two peaks. Based on these data and the results at $E=1$ meV (Fig. 2), we plot in Fig. 1f the dispersions of spin excitations in the normal and SC
states. It is immediately clear that spin excitations of FeSe$_{0.4}$Te$_{0.6}$ display an hourglass-like dispersion in the normal state. 

\begin{figure}[tbp]
\includegraphics[scale=0.46]{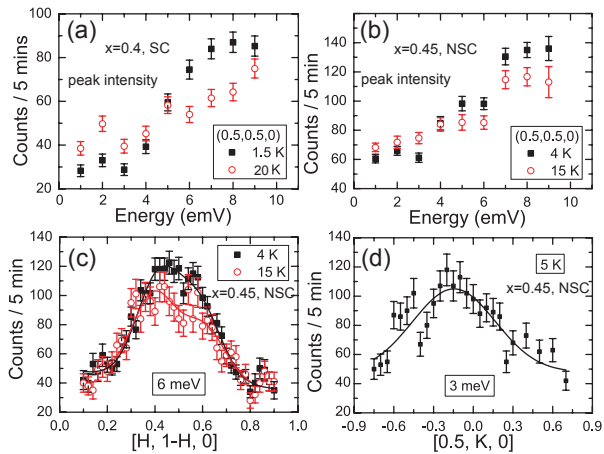}
\caption{ (Color online)
Constant-$Q$ scans at $Q=(0.5,0.5)$ for temperatures above and below $T_c$ in (a) $x=0.4$ and (b) $x=0.45$ samples. (c) Transverse scan at 6 meV in the $x=0.45$ sample that shows weak temperature dependence. (d) Magnetic excitation at 3 meV centered at (0.5,0,0) in the $x=0.45$ sample.
  }
\label{doping}
\end{figure}

To determine the impact of superconductivity on
the energy dependence of spin excitations,
we show in Fig. 4a constant-$Q$ scans at $Q=(0.5,0.5)$ below and above $T_c$ in the $x=0.4$ SC sample. Consistent with earlier work \cite{yqiu,mook},
cooling through $T_c$ re-arranges the scattering profile by reducing the magnetic scattering
below $E\approx 4$ meV and creating a resonance at $E=6.5$ meV (Fig. 4a).

In cuprates,the resonance and hourglass dispersion may arise from the $d$-wave nature of the superconducting gap \cite{norman,eschrig}.
To see if the hourglass dispersion in Fig. 1f is directly connected with superconductivity, we especially prepared a nearly NSC $x=0.45$ sample and carried out identical measurements as those for the SC $x=0.4$ sample.  Although the NSC $x=0.45$
 exhibits an hourglass dispersion remarkably similar to that in the $x=0.4$ SC sample (Figs. 1f and 1g), the incommensurate scattering
at 1 meV show no temperature dependence (Fig. 2d).  In addition, the system has no spin gap and shows only weak resonance above 5 meV (Fig. 4b).
 The transverse scans at 6 meV in Fig. 4c further demonstrate the weak resonance.  Figure 4d shows that there is a large broad peak along $K$-direction at (0.5,0,0), which suggests that our $x=0.45$ sample is indeed poorly SC \cite{zxu}.

The hourglass dispersion may be understood within a Fermi surface nesting picture similar to the case of pure chromium \cite{endoh,fishman}, where the incommensurate spin density waves come from the interband nesting between the electron and hole bands. In the FeSe$_x$Te$_{1-x}$ system, if we consider two nesting wavevectors in a multi-band system, where the two hole pockets ($\alpha2$ and $\alpha3$) at $\Gamma$ point
are nested to the electron pockets at $M$ point as shown in Figs. 1c-1e,
the low energy excitations which disperse inward to the commensurate wavevector $Q=(0.5,0.5)$ start at a incommensurate wavevector about $Q=(0.5-0.15,0.5+0.15)$, which is different from the incommensurate scattering at $Q=(0.5-0.09,0.5+0.09)$ defined by the high energy dispersion \cite{lumsden1,argyriou}. The reason why such nesting conditions are favored may be related to orbital characters\cite{lee,tamai,fchen}.
Although this nesting scheme is appealing, extremely flat band is required to produce the flat dispersion of spin excitations observed at low energies. The bare dispersion calculated by local-density approximation (LDA) \cite{argyriou} is not flat enough to produce the observed excitations. However, we note that recent angle resolved photoemission data \cite{tamai,fchen} have shown that compared with LDA calculations, certain bands in the FeSe$_{x}$Te$_{1-x}$ have very large renormalization factors and become much flatter, possibly due to strong electron-electron correlation effects \cite{si}.  These ARPES results can provide  an consistent understanding of our experimental results based on the Fermi surface nesting between $\alpha3$ and $\beta$ bands as shown in Fig. 1.  Based on the proposed nesting picture in Fig. 1c-1d, we can derive the slope of the lower-energy spin excitation in Fig. 1f-1g as $dE/d\delta \approx 2Qcos^2\theta(1/v_{F,\alpha3}+1/v_{F,\beta})^{-1}$, where $\theta$ is the angle between the incommensurate wavevector $Q$ =(0.5-$\delta$, 0.5+$\delta$) and the AF wavevector (0.5,0.5). We obtain $dE/d\delta \approx$ 50 meV/r.l.u., fitting to our experimental data.  From our experimental value of the slope,  if the two Fermi velocities of  $\alpha3$ and $\beta$ bands are close to each other as shown in \cite{tamai}, we obtain the Fermi velocities to be 45 meV{\AA} for both bands. On the other hand, if the dispersion is predominantly caused by one of the two bands as shown in Ref.\cite{fchen}, where  the renormalization factors around $M$ point are smaller and the $\alpha3$ band is much flatter, we obtain  the velocity of $\alpha3$, $v_{F,\alpha3} \approx$ 22 meV{\AA}.

In summary,  we observe the hourglass dispersion in both SC and NSC FeSe${_x}$Te${_{1-x}}$ samples, phenomenologically similar to the case of cuprates and pure chromium metals. Whether the hourglass behavior of spin excitations is a very common feature in the metallic states of these magnetically-fluctuating systems needs to be further investigated both experimentally and theoretically.

This work is supported by Chinese Academy of Science, 973 Program (2010CB833102), and by the
US DOE, BES, through DOE DE-FG02-05ER46202 and Division of Scientific User
Facilities.

\end{document}